\documentclass{ws-ijmpcs}
\usepackage{slashed}
\usepackage{multirow}
\usepackage[table]{xcolor}

\newcommand{\beqn}{\begin{eqnarray}}
\newcommand{\eeqn}{\end{eqnarray}}
\newcommand{\beqs}{\begin{subequations}}
\newcommand{\eeqs}{\end{subequations}}
\newcommand{\eq}[1]{(\ref{#1})}

\newcommand{\cM}{{\cal M}}

\newcommand{\odd}{\cellcolor[gray]{0.95}}
\newcommand{\nhline}{\hline\\[-3mm]}
\newcommand{\mline}{\nhline\\[-4mm] \odd & \odd & \odd \\[-3mm]}
\begin{document}

\markboth{M. N. Chernodub}
{Vacuum Superconductivity, Conventional Superconductivity and Schwinger Pair Production}

%
\catchline{}{}{}{}{}
%

\title{VACUUM SUPERCONDUCTIVITY, CONVENTIONAL SUPERCONDUCTIVITY AND SCHWINGER PAIR PRODUCTION\footnote{Plenary talk at Quantum Field Theory Under the Influence of External Conditions 2011 (QFEXT11), Benasque, Spain, September 18-24, 2011}}

\author{M. N. CHERNODUB\footnote{On leave from Institute for Theoretical and Experimental Physics, Moscow, Russia. 
\newline 
This work was supported by Grant No. ANR-10-JCJC-0408 HYPERMAG.}}
\address{CNRS, Laboratoire de Math\'ematiques et Physique Th\'eorique,
Universit\'e Fran\c{c}ois-Rabelais, \\ F\'ed\'eration Denis Poisson - CNRS,
Parc de Grandmont, Universit\'e de Tours, 37200 France\\[1mm]
Department of Physics and Astronomy, University of Gent, S9, B-9000 Gent, Belgium\\
E-mail: maxim.chernodub@lmpt.univ-tours.fr}

\maketitle


\begin{abstract}
In a background of a very strong magnetic field a quantum vacuum may turn into a new phase characterized by anisotropic electromagnetic superconductivity. The phase transition should take place at a critical magnetic field of the hadronic strength ($B_c \approx 10^{16}$ Tesla or  $e B_c \approx 0.6\,\mbox{GeV}^2$). The transition occurs due to an interplay between electromagnetic and strong interactions: virtual quark--antiquark pairs popup from the vacuum and create -- due to the presence of the intense magnetic field -- electrically charged and electrically neutral \mbox{spin-1} condensates with quantum numbers of $\rho$ mesons. The ground state of the new phase is a complicated honeycomblike superposition of superconductor and superfluid vortex lattices surrounded by overlapping charged and neutral condensates. In this talk we discuss similarities and differences between the superconducting state of vacuum and conventional superconductivity, and between the magnetic--field--induced vacuum superconductivity and electric--field--induced Schwinger pair production.
\end{abstract}

\ccode{PACS numbers: 12.38.-t, 13.40.-f, 74.90.+n}

\section{Introduction}

Recently, we have suggested that the vacuum in a sufficiently strong magnetic field background may undergo a spontaneous transition to an electromagnetically superconducting phase\cite{Chernodub:2010qx,Chernodub:2011mc}. This unusual effect emerges due to an interplay between strong (gluon-mediated) forces and electromagnetic (photon-mediated) interactions of quarks with the external magnetic field. The electromagnetic and strong interactions are coupled to each other via quarks because the quarks are electrically charged particles which carry a color charge. The electric charge allows for the quark to interact electromagnetically. The color charge is responsible for strong fundamental interaction which, in particular, binds the quarks into hadrons. 

Usually,  electromagnetic interactions play a negligible role in hadronic physics because the electromagnetic interaction (with the coupling $\alpha_{\mathrm{e.m.}} \approx 1/137$) is much weaker compared to the strength of the strongly interacting sector governed by the Quantum Chromodynamics (QCD) with the strong coupling $\alpha_{\mathrm{QCD}} \sim 1$. However, it looks natural that electromagnetic properties of quarks may start to play an important role if the system is subjected to a strong  external magnetic field $B$ with $e B \gtrsim \Lambda_{\mathrm{QCD}}^2$, where $\Lambda_{\mathrm{QCD}} \sim 100\, \mathrm{MeV}$ is a typical massive scale in QCD.

Despite enormous values of the magnetic strength, $e B \gtrsim{(100\, \mathrm{MeV})}^2$, the interest in this topic is not purely academic. For example, the chiral magnetic effect\cite{Fukushima:2008xe} (generation of an {\it electric current}  along the axis of a background {\it magnetic field} in a chirally--imbalanced matter\cite{Vilenkin:1980fu}) should take place in hot quark matter that emerges for a short time in an expanding fireball created in a noncentral collision of heavy ions\cite{Kharzeev:2004ey} because such collisions generate extremely high magnetic fields\cite{Fukushima:2008xe,Skokov:2009qp}. The heavy--ion experiments are currently underway at the Relativistic Heavy Ion Collider  at Brookhaven National Laboratory in USA, and the Large Hadron Collider near Geneva, Switzerland. Moreover, strong magnetic field may also have existed in the early Universe\cite{Grasso:2000wj}, and this strong-field epoch of our history may have also left some traces in the present-day Universe\cite{Chernodub:2010qx,Smolyaninov:2011wc}.

Sufficiently strong magnetic field may affect both matter and the quantum vacuum. A relevant example of the QCD scale is the so-called magnetic catalysis\cite{Klimenko:1991he} which leads to an enhancement of a violation of the chiral symmetry due to increasing external magnetic field. The well-known finite-temperature phase transitions (related to quark liberation and to the chiral restoration) are influenced by the strong magnetic field both in strongly interacting vacuum\cite{Gatto:2010pt} and in strongly interacting matter\cite{Preis:2011sp}. 

We would like to argue that there exists a new phase transition which may take place at relatively low temperatures (in QCD scale) and at a sufficiently strong magnetic field. This transition turns the vacuum into a electromagnetic superconductor\cite{Chernodub:2010qx,Chernodub:2011mc}. The superconductivity of, basically, empty space is caused by a spontaneous creation of (charged) $\rho$-meson condensates out of virtual quarks and antiquarks if the magnetic field becomes stronger than the critical value
\beqn
B_c \simeq 10^{16} \, {\mathrm{Tesla}}
\qquad  {\mathrm{or}} \qquad
e B_c \simeq 0.6\,\mbox{GeV}^2\,.
\quad
\label{eq:Bc}Ê
\eeqn 
The charged $\rho$ mesons is a composite particle which is made of a light quark and a light antiquark. There are positively and negatively charged $\rho$ mesons: $\rho^+ = u \bar{d}$ and $\rho^- = d \bar{u}$, where $u$ and $d$ are up and down quarks, respectively. The electromagnetic superconductivity of the vacuum should be accompanied by condensation of the neutral $\rho$ mesons\cite{Chernodub:2010qx}. The latter effect can be interpreted as a superfluidity\cite{Chernodub:2011tv,ref:Jos}. 

The vacuum superconductivity was found both in an effective bosonic model which describes the electrodynamics of the $\rho$ mesons\cite{Chernodub:2010qx} and in a Nambu--Jona-Lasinio model which describes dynamics of quark degrees of freedom\cite{Chernodub:2011mc}. Signatures of this counterintuitive effect were also found in holographic effective theories\cite{Callebaut:2011ab} and in numerical (``lattice'') approaches to QCD\cite{Braguta:2011hq}. 

\section{The mechanism, qualitatively}

In this section we discuss very basic features of the conventional superconductivity and then turn our attention to its vacuum counterpart.

\subsection{Conventional superconductivity}

In a simplified picture, an electron in a metal behaves as an almost free negatively charged particle which moves in a background of a lattice of positively charged ions. A conventional superconductivity is a result of condensation of specific ``Cooper pairs'' made of some of these electrons. Each Cooper pair can be regarded as a state of two electrons which are loosely bounded together by a small attractive force. The attraction is mediated by a phonon exchange, and the phonon is a quantum of vibration of the ionic lattice.

A simplified picture of the phonon exchange is as follows. Imagine, that an electron moves through the ionic lattice and attracts neighboring ions due to the Coulomb interaction. The local deformation of the ionic lattice leads a local excess of the positive electric charge in a vicinity of the electron. The excess of the positive charge attracts another electron, so that the like-charged electrons may experience a mutual attractive force in a background of the positively charged ion lattice. The lattice distortions can be described as a superposition of collective excitations of the ion lattice (phonons), so that the whole process of the attractive electron-electron interaction can be viewed as a phonon exchange.

The attractive phonon interaction between electrons is so weak that thermal fluctuations may easily wash out the bounding effect of this attractive force. Nevertheless, if the system is cooled down enough, then the attractive interaction prevails the thermal disorder and, consequently, the Cooper pairs may indeed be formed. The formation of the Cooper pairs is guaranteed by the fact that at low temperature the dynamics of the electrons is basically one dimensional, while in one spatial dimension an arbitrarily weak attraction should always lead to appearance of a bound state (the Cooper theorem). The effective dimensional reduction of the electron dynamics from three spatial dimensions to one spatial dimension occurs because at low temperature the interaction between the electrons is possible if and only if (the momenta of) the electrons lie sufficiently close to the Fermi surface. The Cooper pair is formed by two electrons with mutually opposite momenta and spins.

The electrons themselves cannot condense because of the their fermionic nature. However, the Cooper pairs behave effectively as bosons, so that they may appear in a form of a condensate. The condensate of the Cooper pairs is a spatially large (infrared) structure, in which all Cooper pairs behave as one collective entity. The condensate can move frictionlessly through the ion lattice in a manner of a superfluid. Since the Cooper pairs are electrically charged states, their condensate is characterized by infinite conductivity, or, in other words, by zero resistance. Therefore, the condensation of the Cooper pairs make the material superconducting.

Thus, there are three basic ingredients of the conventional superconductivity:
\begin{itemize}
\item[A)] the presence of electric charge carriers in the material (i.e., electrons in a metal or alloy);
\label{page:condition:A}
\item[B)] the reduction of the dynamics of the electric charge carriers from three spatial dimensions to one spatial dimension; 
\label{page:condition:B}
\item[C)] an attractive interaction between the charge carriers (which are like-charged particles).
\label{page:condition:C}
\end{itemize}

How these ingredients may appear in the vacuum superconductivity?

\subsection{Vacuum superconductivity}

\subsubsection{Environment}
\label{ref:environment}

Contrary to the conventional superconductivity, there are no charge carriers in the vacuum in its low--temperature phase. The vacuum is obviously an insulator. However, due to quantum fluctuations the vacuum can be regarded as a boiling soup of virtual particles. Some of these virtual particles may convert to real particles and lead to unusual transport phenomena in certain cases. 

Indeed, a transition from the boiling soup of virtual particles to the real world is not forbidden {\it ab initio}. There are at least two  relevant examples when such ``virtual-to-real'' transition. The first example is the Schwinger effect: in a uniform time-independent background of a sufficiently strong external electric field the vacuum should produce potentially detectable electron-positron pairs\cite{ref:Schwinger,ref:QED:strong}${}^,$\footnote{An analogy of the magnetic-field-induced vacuum superconductivity and the Schwinger effect will be discussed later in Section~\ref{sec:Schwinger}.}. 

The second example is related to thermal fluctuations: a thermal bath may turn the vacuum from the insulating regime to a conducting phase at critical temperature 
\beqn
T^{\mathrm{QED}} \approx 2 m_e \approx 1\,\mbox{MeV} \approx 10^{10}\,\mathrm{K}\,,
\eeqn 
which is of the order of QED (Quantum Electrodynamics) scale. Indeed, at $T \sim 0.1 \, T^{\mathrm{QED}}$ the vacuum gets ionized to form a neutral electron-positron plasma because of thermal activation of the virtual $e^+e^-$ pairs. The QCD sector of the vacuum contributes to conductivity at much higher temperatures $T \sim T_c^{\mathrm{QCD}}$ where
\beqn
T_c^{\mathrm{QCD}} \approx 170\,\mbox{MeV} \approx 2 \times 10^{12}\,\mathrm{K}\,,
\eeqn 
is the critical temperature of the quark liberation (``de-confinement'') crossover transition to the quark-gluon plasma state\cite{ref:Zoltan}. 

Thus, in certain external conditions (strong electric field, high temperatures) the quantum vacuum may become conducting. And, as we will argue below, in strong magnetic field the vacuum becomes {\underline{super}}conducting (with zero electric resistance). In the magnetic--field--induced vacuum superconductivity the key role is played by virtual quarks and antiquarks which have fractional electric charges\footnote{Due to the quark confinement phenomenon realized at relatively ``low'' temperatures, $T<T_c^{\mathrm{QCD}}$, the quarks and anti-quarks appear always in a form quark-antiquark states (mesons) or three-quark states (baryons) which always have integer valued electric charges.}.

Basic properties of the underlying environments of a conventional superconductor (a metal) and the quantum vacuum are summarized\footnote{We do not mention virtual electrons and virtual positrons in Table~\ref{tbl:environment} because they do not play any substantial role in this exotic superconductivity mechanism.} in Table~\ref{tbl:environment}.

\begin{table}[ph]
\tbl{Superconductivity: conventional vs. vacuum (environment)}
{\begin{tabular}{lll}
\mline
\multicolumn{3}{c}{\odd environment}\\[1mm]
\nhline
 & conventional & vacuum \\[1mm]
\nhline
nature of carriers &  real & virtual\\[1mm]
\nhline
realized in &  a material (metal, alloy etc) &   vacuum (empty space)\\[1mm]
\nhline
under usual conditions &  a conductor &  an insulator\\[1mm]
\nhline
basic carriers  of & \multirow{2}{*}{electrons ($e$)} & quarks ($u$, $d$) and \\[1mm]
electric charge &  & antiquarks  ($\bar u$, $\bar d$) \\[1mm]
\nhline
electric charges    &  \multirow{2}{*}{ $q_e = - e$ \quad ($e \equiv |e|$)} &  $q_u = + 2 e/3$,\ \ $q_d = - e/3$  \\[1mm]
of basic carries     &  &  $q_{\bar u} = - 2 e/3$,\ \  $q_{\bar d} = + e/3$  \\[1mm]
\nhline
\end{tabular}
\label{tbl:environment}}
\end{table}

Thus, the condition "A" of Section~\ref{ref:environment} (the presence of electric charge carriers) is (virtually) satisfied in the quantum vacuum.

\subsubsection{Superconducting carriers}

Condition ``B'' of superconductivity (Section~\ref{ref:environment}, page~\pageref{page:condition:B}) states that one needs a dimensional reduction of the fermion dynamics in order for the superconducting carriers to be created. In a conventional superconductor it is the Fermi surface that leads to the effective dimensional reduction of the electron dynamics and facilitates the formation of the Cooper pairs. On the contrary, in the vacuum all chemical potentials are zero because of the absence of matter, and the Fermi surfaces do not exist.

However, the role of the Fermi surface may be played by the magnetic field: in a background of a sufficiently strong  magnetic field the dimensional reduction of low-energy charges does indeed occur since electrically charged particles can move only along the lines of the magnetic field. This effect leads to the required dimensional reduction from three to one spatial dimensions.

The dimensional reduction effect works for all electrically charged elementary particles, including electrons, positrons, quarks, antiquarks etc. Notice, however, that for the formation of the electrically charged bound states (analogues of the Cooper pairs) one should require the following:
\begin{itemize}
\item[(i)] the constituents of the superconducting bound states should be likewise charged in order to guarantee that their bound state is electrically charged;
\item[(ii)] the interaction between these constituents should be attractive. 
\end{itemize}

The requirements (i) and (ii) rule out a potential importance of the pure QED vacuum sector which describes electrons, positrons and photons. Indeed, in the vacuum the electron-electron interaction is repulsive due to the photon exchange so that analogues of the Cooper pairs cannot be created from the virtual electrons. Despite of the fact that the photon-mediated interaction between an electron and a positron is attractive,  the electron--positron bound state is not interesting for our purposes because this bound state~is 
\begin{itemize}
\item[(a)] electrically neutral; 
\item[(b)] unstable due to fast annihilation processes. 
\end{itemize}
Thus, the vacuum superconductivity cannot emerge in the pure QED vacuum.

Therefore, if electrically charged bound states are formed in the strong magnetic field they should be outside of the pure QED sector of the vacuum. And, indeed, we notice that the strongly interacting (QCD) sector of the vacuum does contain an analogue of the phonon which may attract the-like electrically charged particles (condition "C" of superconductivity, Section~\ref{ref:environment}, page~\pageref{page:condition:C}). The ``vacuum'' analogue of the phonon is a gluon which is a carrier of the strong force. The gluon provides an attractive interaction between the quarks and the anti-quarks and binds them in the pairs called mesons. The quarks and antiquarks are always  bound by the gluon exchange regardless of the sign of their electric charges (one can consult Table~\ref{tbl:environment}). For example,  a $u$ quark with the electric charge $q_u = + 2 e/3$ and a $\bar d$ antiquark with the electric charge $q_{\bar d} \equiv - q_d = + e/3$ are bound by the gluon-mediated interaction, forming the $\rho$ meson with the electric charge $q_\rho = +e$. Therefore, the vacuum analogue of the Cooper pair is the charged $\rho$ meson state. And in next sections we show that the $\rho$--meson condensates do indeed appear in the vacuum in background of the strong magnetic field, and we argue that the emergent state is indeed a superconductor.

The vacuum superconductor has an exotic property, a strong anisotropy of the superconductivity. Indeed, the key requirement  of our mechanism is the dimensional reduction (condition "B" of Section~\ref{ref:environment}, page~\pageref{page:condition:B}) which states that the constituent electric charges (the quarks $u$ and $b$ and their antiquarks) can move only along the axis of the magnetic field. Therefore, the superconducting charge carriers (the $\rho$ mesons in our case) can flow only along the axis of the magnetic field. Thus, the vacuum is superconducting in the longitudinal direction (along the magnetic field) while in two transverse directions the vacuum behaves as insulator. Due to the anisotropic superconductivity the vacuum in strong magnetic field becomes a (hyperbolic) metamaterial which, electromagnetically, behaves as diffractionless ``perfect lenses''\cite{Smolyaninov:2011wc}.

In Table~\ref{tbl:carriers} we provide a comparison of the basic features of the superconducting carriers in the framework of the standard conventional superconductivity and the exotic vacuum superconductivity.

\clearpage
\begin{table}[ph]
\tbl{Superconductivity: conventional vs. vacuum (superconducting carriers)}
{\begin{tabular}{lll}
\mline
\multicolumn{3}{c}{\odd superconducting carriers (SCC)}\\[1mm]
\nhline
SCC type:  & Cooper pair & two $\rho$--meson excitations, $\rho^\pm$\\[1mm]
\nhline
\multirow{2}{*}{composition of SCC:}   & \multirow{2}{*}{electron-electron state ($ee$)} & quark-antiquark states\\[1mm]
&   & ($\rho^+ = u\bar d$ and $\rho^- = d\bar u$) \\[1mm]
\nhline
electric charge of SCC:  & $- 2 e$ & $+ e$ and $-e$, respectively\\[1mm]
\nhline
spin of SCC:  & typically zero (scalar) & one (vector) \\[1mm]
\nhline
                                                            & \multicolumn{2}{l}{1) reduction of dynamics of basic electric charges}\\[1mm]
   the SCCs are                  & \multicolumn{2}{l}{\phantom{1) }from three spatial dimensions to one dimension}	\\[1mm]
\cline{2-3}\\[-3mm]   
   formed due to                               & 2) attraction force                              &  2) attraction force between \\[1mm]
                                                            & \phantom{2) }between two electrons &  \phantom{2) }a quark and an antiquark \\[1mm]
\nhline
\multirow{3}{*}{
\begin{tabular}{l}
\hskip 0mm 1) a reason for the \\[1mm] 
\hskip 0mm reduction $3d \to 1d$
\end{tabular}} 
 & at very low temperatures &  in strong magnetic field the\\[1mm]
 &  electrons interact with each  &  motion of electrically charged \\[1mm]
& other near the Fermi surface  & particles is one dimensional \\[1mm]
\cline{2-3}\\[-3mm]
\hskip 3mm 2) attraction is due to & phonons (lattice vibrations) & gluons (strong force, QCD)\\[1mm]
\nhline
anisotropy of & \multirow{3}{*}{\begin{tabular}{c} superconducting \\[1mm] in all directions \end{tabular}} & superconducting only along \\[1mm]
superconducting &  & the axis of the magnetic field,\\[1mm]
properties &  & insulator in other directions\\[1mm]
\nhline
\end{tabular}
\label{tbl:carriers}}
\end{table}

Summarizing, in strong magnetic field the dynamics of virtual quarks and antiquarks is effectively one-dimensional because these electrically charged particles tend to move along the lines of the magnetic field. In one spatial dimension a gluon-mediated attraction between a quark and an antiquark inevitably leads to formation of a quark-antiquark bound state. This bound state is composed of quarks of different flavors ($u \bar{d}$ and $d \bar u$) which allows for the bound state (i) to be electrically charged and (ii) gain stability agains direct annihilations.  Moreover, it turns out that the bound state should be a vector (spin-triplet) state in order to occupy a lowest energy state. This superconducting bound state has quantum numbers of an electrically charged $\rho^\pm$ meson.

\subsubsection{Meissner effect and magnetic field}

Suppose, our qualitative considerations are correct and, indeed, in the background of strong enough magnetic field the vacuum turns into a superconducting state. Then, we immediately come to two would-be controversies (Table~\ref{tbl:magnetic}):
\begin{itemize}
\item[a)] the vacuum superconductivity is induced by the strong magnetic field while the all existing superconductors are known to be destroyed by a strong enough magnetic field;
\item[b)] the existence of the superconducting phase in the magnetic field background indicates that the vacuum in the new superconducting phase does not screen the magnetic field in the bulk while all known superconductors expel/screen weak external magnetic field (the Meissner effect).
\end{itemize}

\begin{table}[ph]
\tbl{Superconductivity: conventional vs. vacuum (magnetic field and temperature)}
{\begin{tabular}{lll}
\mline
\multicolumn{3}{c}{\odd magnetic field, thermal fluctuations and superconductivity}\\[1mm]
\nhline
role of the magnetic field & destroys superconductivity & enhances superconductivity \\[1mm]
\nhline
the Meissner effect & present & absent \\[1mm] 
\nhline
temperature & destroys superconductivity & destroys superconductivity \\[1mm] 
\hline
\end{tabular}
\label{tbl:magnetic}}
\end{table}

A common resolution of these two puzzling issues is rather simple. Qualitatively, the conventional Meissner effect is caused by large superconducting currents which are induced by the external magnetic  field in the bulk of the superconductor. The circulation of these currents in the transversal (with respect to the magnetic field axis) plane generates a backreacting magnetic field which tend to screen the external magnetic field in the bulk of the superconducting material. Moreover, in strong enough external magnetic field the (positive) excess in energy of the induced transverse currents prevails the (negative) condensation energy. As a result, in a too strong magnetic field background the conventional superconductivity becomes energetically unfavorable and at certain critical field the material turns from the superconducting state back to the normal state. Indeed, it is the negative condensation energy which makes the superconducting state energetically favorable with respect to the normal state and in strong magnetic fields the energy surplus becomes smaller than the energy deficit associated with the superconducting state.

As for the vacuum superconductor, the absence of the true Meissner effect is a natural consequence of the anisotropy of the vacuum superconductivity. The anisotropy comes, as we have already discussed, as a natural consequence of the dimensional reduction of the charge dynamics in the strong magnetic field background. In the transversal directions the vacuum behaves as an insulator, so that no large transversal electric currents can be induced by the external magnetic field in the vacuum superconducting state. In the absence of the transversal superconductivity the external magnetic field cannot be screened by the longitudinally superconducting state of the vacuum. Thus, there is no screening backreaction to the external magnetic field, and, consequently, there is no energy deficit which could make the superconducting state less favorable compared to the usual insulating state. Therefore, the external magnetic field cannot destroy the electromagnetic superconductivity of the vacuum once the superconducting state is created.

\clearpage

\subsubsection{Role of temperature}

The thermal fluctuations should destroy the vacuum superconductivity similarly to the conventional one (Table~\ref{tbl:magnetic}). Indeed, the key element of the vacuum superconductivity is the dimensional reduction of the charge dynamics: the quarks and antiquarks move predominantly along the axis of the magnetic field. Physically, the one--dimensional motion is a natural consequence of the fact that relevant electric charges occupy the lowest Landau level which is localized in the transverse plane. The one-dimensional motion could be destroyed by jumps of the particles to higher Landau levels which are less localized. In a magnetic field of the order of the critical one, $B \sim B_{c}$, a typical difference between the energy levels is of the QCD scale, $\delta E \sim \Lambda_{\mathrm{QCD}} \approx 100\,\mbox{MeV}$. Therefore,  strong enough thermal fluctuations of the  QCD scale, $T \sim \Lambda_{\mathrm{QCD}} \approx 100\,\mbox{MeV}$, should destroy the dimensional reduction property and, consequently, the superconductivity at certain temperature $T_{c} \equiv T_{c}(B)$.

\subsubsection{Welcome to the real world: an analogy with Schwinger effect}
\label{sec:Schwinger}

The Schwinger effect is production of the electron--positron pairs from the vacuum in a background of a strong enough {\emph{electric}} field\cite{ref:Schwinger}: real particles are created out of virtual ones. The created particles tend to screen the external electric field.  

In a similar fashion, the vacuum superconductivity is associated with creation of the charged quark-antiquark condensates out of vacuum subjected to the strong enough {\emph{magnetic}} field\cite{Chernodub:2010qx,Chernodub:2011mc}. In a clear contrast to the Schwinger effect, these quark-antiquark pairs form a condensate which  does not tend to screen the external magnetic field.  Moreover, the energy scales of the processes associated with Schwinger effect (QED scale) and the vacuum superconductivity (QCD scale) are different by three orders of magnitude. The energy difference reflects a difference in the nature of the interactions (QED vs QCD) involved in the corresponding mechanisms.

We outline differences between and similarities of these two effects in Table~\ref{tbl:Schwinger}.

\begin{table}[ph]
\tbl{Vacuum superconductivity vs. Schwinger effect}
{\begin{tabular}{lll}
\nhline
 & Schwinger effect  & Vacuum superconductivity \\[1mm]
\nhline
environment & vacuum &  vacuum \\[1mm]
\nhline
interactions involved & electromagnetic only & electromagnetic and strong \\
\nhline
energy scale & megaelectronvolts (MeV $\equiv 10^6$ eV) & gigaelectronvolts (GeV $\equiv 10^9$ eV) \\
\nhline
background of & strong electric field, $E$ & strong magnetic field, $B$ \\[1mm]
\nhline
\multirow{2}{*}{critical value} & $E_c = m_e^2/e \approx 10^{18}\,\mathrm{V/m}$ & $B_c = m_\rho^2/e \approx 10^{16}\,\mathrm{T} $ \\[1mm]
& {\scriptsize{($m_e = 0.511\,\mbox{MeV}$ is electron mass)}} & {\scriptsize{($m_\rho = 0.775 \,\mbox{GeV}$ is $\rho$--meson mass)}} \\
\nhline
\multirow{2}{*}{nature} & virtual electron-positron ($e^- e^+$)  &  virtual quark-antiquark ($u \bar{u}$, $d \bar{u}$)   \\
  & pairs become real $e^- e^+$ pairs & form real $u \bar{d}$ and $d \bar{u}$ condensates \\
\nhline
\multirow{2}{*}{backreaction} & created $e^- e^+$ pairs tend  &  created $u \bar{d}$ and $d \bar{u}$ condensates  \\
  & to screen the external field & do not screen the external field\\
\nhline
stability & a process (unstable)  & a ground state (stable)  \\
\nhline
\end{tabular}
\label{tbl:Schwinger}}
\end{table}

\subsubsection{Scales and models}
\label{sec:quantities}

The order of magnitude of typical quantities in the conventional superconductors and in its vacuum counterpart are drastically different (Table~\ref{tbl:quantities}). The difference in the critical temperatures $T_c$ at which the corresponding superconducting states are destroyed by thermal fluctuations is about 11 orders of magnitude. The strength of the magnetic field at which the conventional superconductivity is destroyed is about $15\dots17$ orders of magnitude lower compared to the critical magnetic field at which the vacuum turns in the superconducting state. The typical coherence lengths (the size of the Copper pair and the size of the $\rho$--meson, respectively) show difference in 8 orders. Finally, the penetration depth of the vacuum superconductor is infinite.

\begin{table}[ph]
\tbl{Superconductivity: conventional vs. vacuum (basic quantities)}
{\begin{tabular}{lll}
\mline
\multicolumn{3}{c}{\odd order of magnitude of basic quantities}\\[1mm]
\nhline
& & \\[-3mm]
critical temperature & $T_c \sim  10$ K & $T_c \sim 10^{12}$ K ($T_c \sim 100$ MeV)\\[1mm]
\nhline
& & \\[-3mm]
 \multirow{2}{*}{critical magnetic field} &$B_c \sim 0.1$ T (type I)  & \multirow{2}{*}{$B_c \sim 10^{16}$ T ($e B_c \sim 1$ GeV)}\\[1mm]
 & $B_c \sim 10$ T (type II) & \\[1mm]
\nhline
& & \\[-3mm]
size of SCC & \multirow{2}{*}{$\xi \sim 100\ {\mathrm{nm}} =10^{-7}\ {\mathrm{m}}$} & \multirow{2}{*}{$\xi \sim 1\ {\mathrm{fm}} =10^{-15}\ {\mathrm{m}}$} \\[1mm]
(coherence length) & & \\[1mm]
\nhline
& & \\[-3mm]
penetration length & $\lambda \sim 100\ {\mathrm{nm}} =10^{-7}\ {\mathrm{m}}$ & $\infty$\\[1mm]
\nhline
\end{tabular}
\label{tbl:quantities}}
\end{table}

Phenomenologically, both conventional and vacuum superconductivities can be described in the framework of both microscopic fermionic models and macroscopic bosonic theories. The fermionic models describe the basic carriers (electrons or quarks, respectively), and are, generally, nonrenormalizable.  The bosonic models are based on renormalizable Lagrangians which describe the bosonic superconducting excitations (Cooper pairs or $\rho$ mesons, respectively) treating them as pointlike particles. Despite of the fact that fermionic and bosonic approaches have distinct disadvantages, they can describe the superconductivity at a good qualitative (and, often, quantitative) level. In Table~\ref{tbl:models} we outline the correspondence between traditional models of conventional and vacuum superconductivities.

\begin{table}[ph]
\tbl{Superconductivity: conventional vs. vacuum (simplest models)}
{\begin{tabular}{lll}
\mline
\multicolumn{3}{c}{\odd simplest models of superconductivity}\\[1mm]
\nhline
  & conventional & vacuum \\[1mm]
\nhline
\multirow{2}{*}{bosonic} & \multirow{2}{*}{Ginzburg--Landau  model\cite{ref:GL}} & $\rho$-meson electrodynamics\cite{Djukanovic:2005ag},  \\[1mm]
&  & vector dominance model\cite{Sakurai:1960ju} \\[1mm]
\nhline
\multirow{2}{*}{fermionic} &  Bardeen--Cooper-- & Nambu--Jona-Lasinio model\cite{ref:NJL} \\[1mm] 
 & Schrieffer model\cite{ref:BCS} & with vector interactions\cite{Ebert:1985kz} \\[1mm]
\end{tabular}
\label{tbl:models}}
\end{table}

\subsubsection{Back to condensed matter: Reentrant superconductivity}

In most superconductors an increasing external magnetic field suppresses superconductivity via diamagnetic and Pauli pair breaking effects. However, in a very strong magnetic field the Abrikosov flux lattice of a type-II superconductor may enter a quantum limit of the low Landau level dominance\cite{ref:Tesanovic}, characterized by the absence of the Meissner effect, a spin-triplet pairing, and a superconducting flow along the magnetic field axis. Our proposal\cite{Chernodub:2010qx,Chernodub:2011mc} of the vacuum superconductivity has similar features. The suggested quantum limit of the type--II superconductors -- which is sometimes called as the magnetic-field-induced, or ``reentrant'', superconductivity -- is characterized by a very strong (in condensed-matter scale) magnetic field, so that the magnetic length $l_B = \sqrt{2 \pi/|eB|}$ becomes smaller than the correlation length $\xi$.

Although it is unclear whether this particular magnetic-field-induced me\-cha\-nism\cite{ref:Tesanovic} works in real superconductors, the reentrant superconductivity was experimentally observed in certain materials like an uranium superconductor URhGe\cite{ref:Uranium}.

\section{Ground state of vacuum superconductor}

\subsection{Qualitative energy arguments}

It is known that the life of the $\rho$ meson is very short: the lifetime is approximately equal to the time needed for light to pass through the $\rho$-meson itself. However, simple kinematical arguments show that  in a background of sufficiently strong magnetic field the charged $\rho$ meson becomes stable against all known decay modes\cite{Chernodub:2010qx}.

Why  $\rho$ mesons should condense in strong magnetic field? As an illustration, consider a free relativistic spin-$s$ particle moving in a background of the external magnetic field $B$. The energy levels $\varepsilon$ of the  pointlike particle are:
\beqn
\varepsilon_{n,s_z}^2(p_z) = p_z^2+(2 n - g s_z + 1) |eB| + m^2\,, \qquad g = 2\,,
\label{eq:energy:levels}
\eeqn
where the integer $n\geqslant 0$ labels the energy levels, and other quantities characterize the properties of the particle: mass $m$, the projection of the spin $s$ on the field's axis $s_z = -s, \dots, s$, the momentum along the field's axis, $p_z$, and the electric charge $e$. In Eq.~\eq{eq:energy:levels} the gyromagnetic ratio (or, ``$g$--factor'') is taken to be\footnote{This anomalously large value -- which plays an important role in our discussion -- is supported by the renormalizability arguments in an effective $\rho$--meson electrodynamics\cite{Djukanovic:2005ag} outlined in the next Section. The value $g \approx 2$ for the $\rho$ mesons was independently obtained in the framework of the QCD sum rules\cite{ref:g2:sumrules}, in the Dyson--Schwinger approach to QCD\cite{ref:g2:DS}, and, finally, this anomalous value of the $g$--factor is conformed by the first-principle numerical simulations of lattice QCD\cite{ref:g2:lattice}.} $g=2$. 

It is clear from Eq.~\eq{eq:energy:levels} that the ground state corresponds to the quantum numbers $p_z=0$, $n_z = 0$ and $s_z = s$. The ground state energy (or ``mass'', since we consider the state with zero momentum) of the charged $\rho$ mesons ($s=1$) is 
\beqn
m_{\rho^\pm}^2(B) = m_{\rho}^2 - e B\,.
\label{eq:m2:pi:B}
\eeqn
Thus, the mass of the charged $\rho$ meson should decrease with the increase of the magnetic field $B$. When the magnetic field reaches the critical strength~\eq{eq:Bc}Ê, the ground state energy of the $\rho^\pm$ mesons becomes zero.  As the field becomes stronger than the critical value~\eq{eq:Bc}, the ground state energy of the charged $\rho$ mesons gets purely imaginary thus signaling a tachyonic instability of the ground state of the vacuum towards the $\rho$--meson condensation.

The condensation of the $\rho$ mesons in QCD is similar to the Nielsen-Olesen instability of the pure gluonic vacuum in Yang-Mills theory\cite{ref:NO}
(also at finite temperature\cite{Bordag:2006pr}) and to the magnetic-field-induced Ambj\o rn--Olesen condensation of the $W$-bosons in the standard electroweak model\cite{ref:AO}. Both the $\rho$ mesons in QCD, the gluons in Yang-Mills theory, and the $W$ bosons in the electroweak model have the anomalously large $g$--factor, $g \approx 2$.

\subsection{Electrodynamics of $\rho$ mesons}
\label{sec:VDM}

The quantum electrodynamics for the $\rho$ mesons is described by the Lagrangian\cite{Djukanovic:2005ag}:
\beqn
\begin{array}{rcl}
{\delta \cal L} =
- \frac{1}{2} (D_{[\mu,} \rho_{\nu]})^\dagger D^{[\mu,} \rho^{\nu]} + m_\rho^2 \ \rho_\mu^\dagger \rho^{\mu} 
 - \frac{1}{4} \rho^{(0)}_{\mu\nu} \rho^{(0) \mu\nu}{+}\frac{m_\rho^2}{2} \rho_\mu^{(0)}
\rho^{(0) \mu} +\frac{e}{2 g_s} F^{\mu\nu} \rho^{(0)}_{\mu\nu}\,, 
\end{array}
\ \
\label{eq:L} 
\eeqn
where $D_\mu = \partial_\mu + i g_s \rho^{(0)}_\mu - ie A_\mu$ is the covariant derivative,
$g_s \equiv g_{\rho\pi\pi} \approx 5.88$ is the $\rho\pi\pi$ coupling,
$A_\mu$ is the photon field with the field strength $F_{\mu\nu} = \partial_{[\mu,} A_{\nu]}$, the fields
$\rho_\mu = (\rho^{(1)}_\mu - i \rho^{(2)}_\mu)/\sqrt{2}$ and $\rho^{(0)}_\mu \equiv \rho^{(3)}_\mu$
correspond, respectively, to the charged and neutral vector mesons with the mass~$m_\rho$, 
and $\rho^{(0)}_{\mu\nu} = \partial_{[\mu,} \rho^{(0)}_{\nu]} - i g_s \rho^\dagger_{[\mu,} \rho_{\nu]}$.
The model possesses the electromagnetic $U(1)$ gauge invariance: $\rho_\mu(x) \to e^{i \omega(x)} \rho_\mu(x)$, 
$A_\mu(x) \to A_\mu(x) + \partial_\mu \omega(x)$. The last term in Eq.~\eq{eq:L} describes a nonminimal 
coupling of the $\rho$ mesons to electromagnetism implying the $g = 2$ value of the  $g$--factor. 

The quadratic part of the energy density of~\eq{eq:L} in a magnetic field background,
\beqn
\epsilon_0^{(2)}(\rho_\mu) = \sum_{i,j=1}^2\rho_i^\dagger \cM_{ij} \rho_j + m_\rho^2 (\rho_0^\dagger \rho_0 + \rho_3^\dagger \rho_3)\,, \qquad\quad
\cM = 
\left(
\begin{array}{cc}
m_\rho^2 & i e B \\
- i e B & m_\rho^2
\end{array}
\right)\,,
\label{eq:cM}
\eeqn
shows that the mass terms for $\rho_0$ and $\rho_3$ components are diagonal and their prefactors $m_\rho^2$ are unaltered by the external magnetic field. However, the Lorentz components $\rho_1$ and $\rho_2$ are mixed by the non--diagonal mass matrix $\cM$ with the following eigenvalues $\mu_{\pm}^2 = m_\rho^2 \pm e B$ and eigenvectors $\rho_{\pm} = (\rho_1 \mp i \rho_2) / \sqrt{2}$.

The $\rho_-$ state with the mass~\eq{eq:m2:pi:B} becomes unstable and condenses, $\langle \rho_- \rangle \neq 0$, if the magnetic field exceeds the critical value~\eq{eq:Bc}. The condensation of the electrically charged particles leads to the electromagnetic superconductivity\cite{Chernodub:2010qx}. The condensation of the neutral mesons may imply, in turn,  superfluidity of the ground state\cite{Chernodub:2011tv,ref:Jos}. The emerging electrically charged ($\rho$) and electrically neutral ($\rho^{(0)}$) condensates are nontrivial functions of the transverse coordinates $x_1$ and $x_2$\footnote{In a dense isospin--asymmetric matter the longitudinal condensates ($\rho_{0,3} \neq 0$) emerge\cite{ref:isospin} and they lead to an electromagnetically superconducting state\cite{Ammon:2008fc}. We demonstrate the existence of the superconductivity in the vacuum (with transversal condensates, $\rho_{1,2} \neq 0$) instead of dense matter.},
\beqn
\begin{array}{rcl}
\mbox{charged condensate:} \hskip 2.5mm \qquad \rho_1 & = &- i \rho_2 = \rho\,, \qquad \hskip 5mm \rho_0 = \rho_3 = 0\,, \\
\mbox{neutral condensate:} \qquad \rho^{(0)} & = & \rho^{(0)}_1 + i \rho^{(0)}_2\,, \qquad \rho^{(0)}_0 = \rho^{(0)}_3 = 0\,.
\end{array}
\eeqn

\subsection{Periodic structure of the ground state}

\subsubsection{Charged and neutral condensates}
\begin{figure}
\begin{center}
\begin{tabular}{ll}
\includegraphics[width=57mm, angle=0]{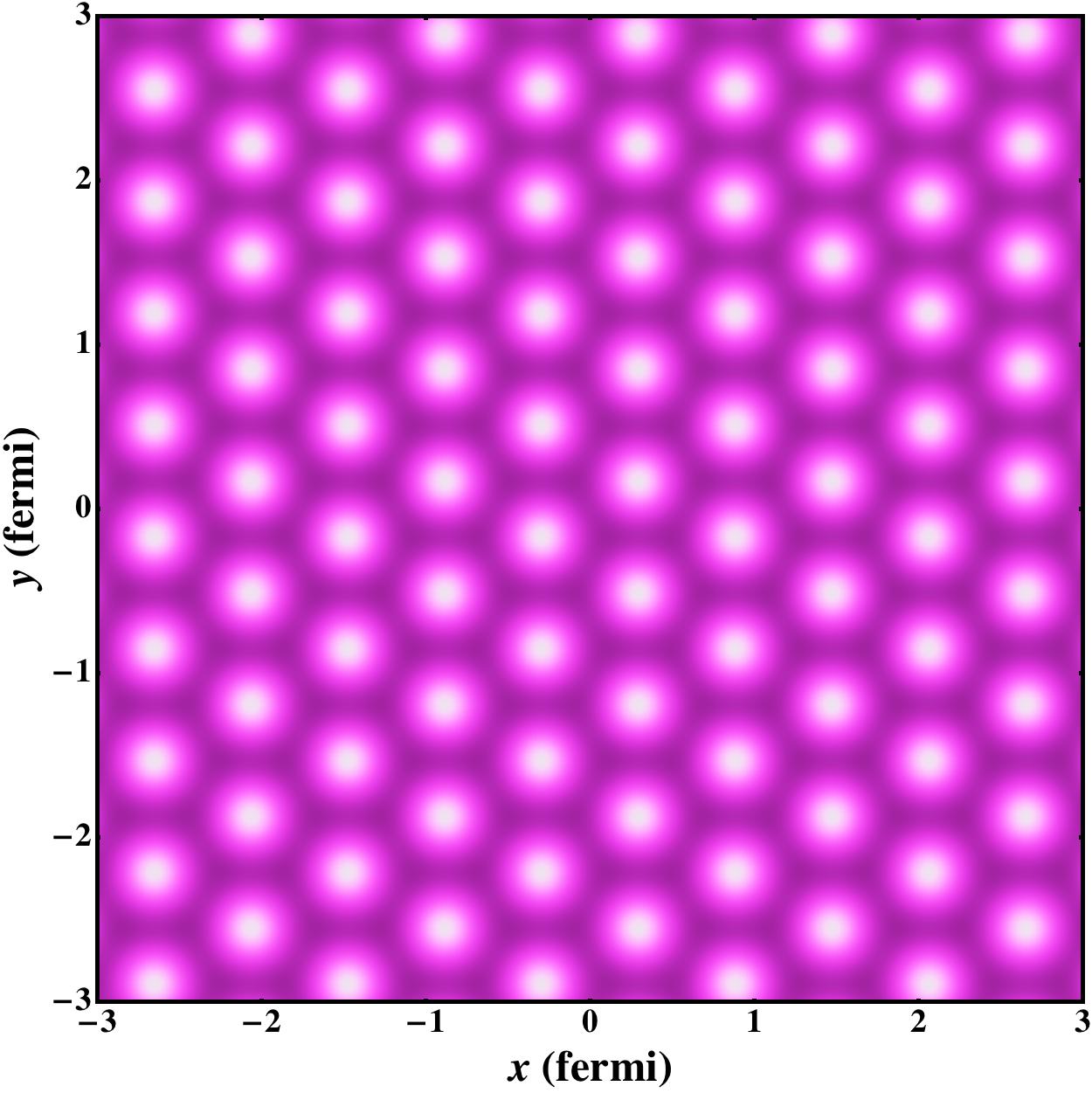} 
& \hskip 5mm
\includegraphics[width=57mm, angle=0]{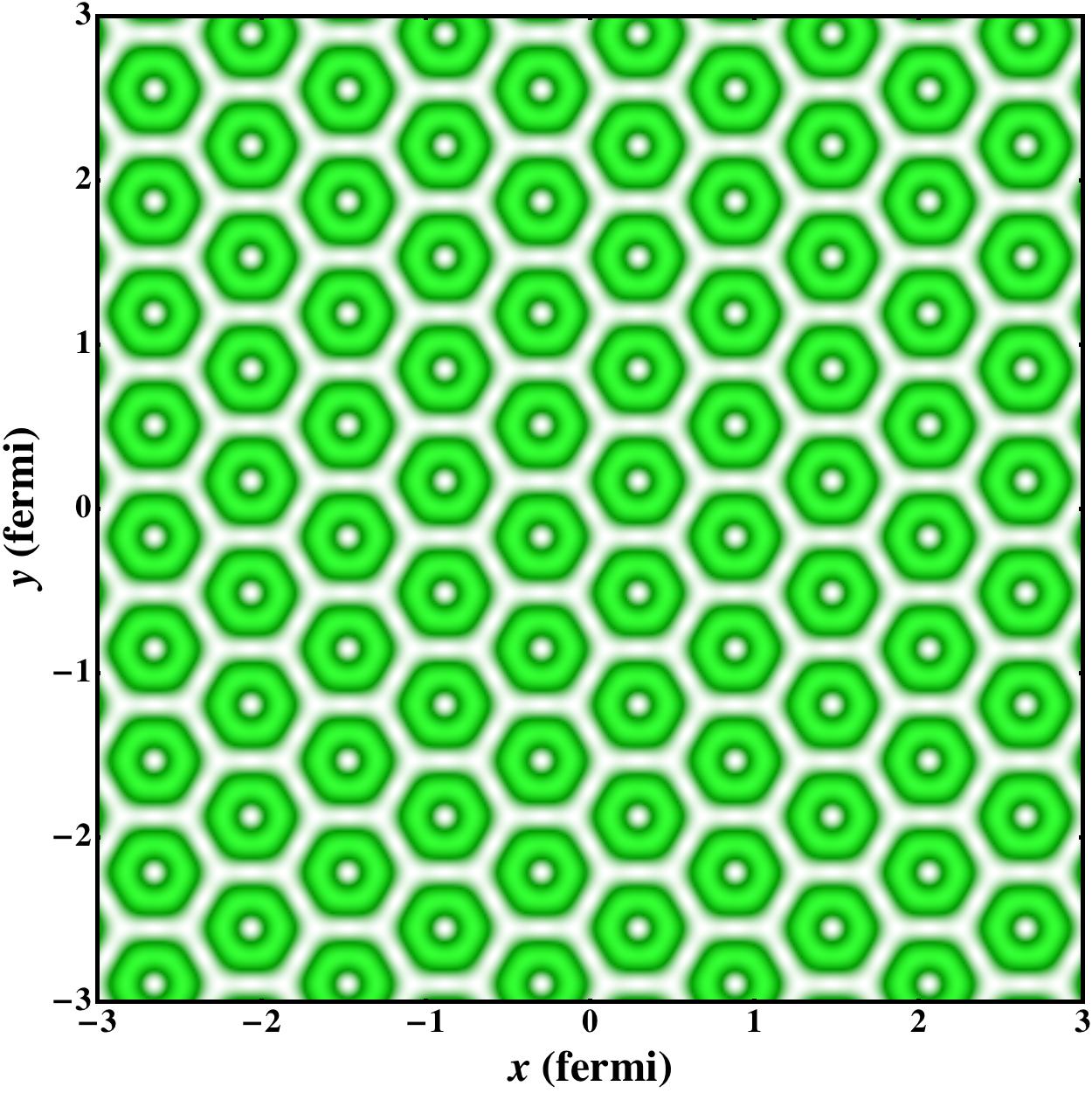}
\end{tabular}
\end{center}
\caption{The density plots of the absolute value of (left) the charged (superconducting) $\rho$--meson condensate $\rho(x_1,x_2)$ and (right) the neutral (superfluid) $\rho$--meson condensate $\rho^{(0)}(x_1,x_2)$ in the transversal $(x_1,x_2)$ plane in the background magnetic field $B=1.01\,B_c$ (Ref.~[29]).}
\label{fig:absolute}
\end{figure}

In the transversal plane the charged $\rho$--meson condensate has a typical periodic Abrikosov pattern\cite{Abrikosov:1956sx}, Figure~\ref{fig:absolute}(left), exposing an equilateral triangle lattice\cite{ref:Jos}. The structure of the neutral condensate is more complex, Figure~\ref{fig:absolute}(right). 

One should also mention that the exotic superconducting phase has all global quantum numbers of the vacuum. For example, all chemical potentials in the superconducting phase are zero. The vacuum is electrically neutral: the presence of the positively charged condensate $\rho$  implies automatically the appearance of a negatively charged condensate $\rho^*$ of equal magnitude ($\rho \equiv |\rho^*|$).  As a result, the energy of the vacuum is lowered, while the net electric charge of the vacuum stays zero\cite{Chernodub:2010qx,Chernodub:2011mc}. Despite of the net electric neutrality, the vacuum may superconduct since a weak external electric field  -- if it is directed along the magnetic field axis -- pushes the positively and negatively charged condensates in opposite directions, thus creating a net electric current of double magnitude. 

\subsubsection{Superconductor and superfluid vortices}

\begin{figure}
\begin{center}
\begin{tabular}{ll}
\includegraphics[width=56mm, angle=0]{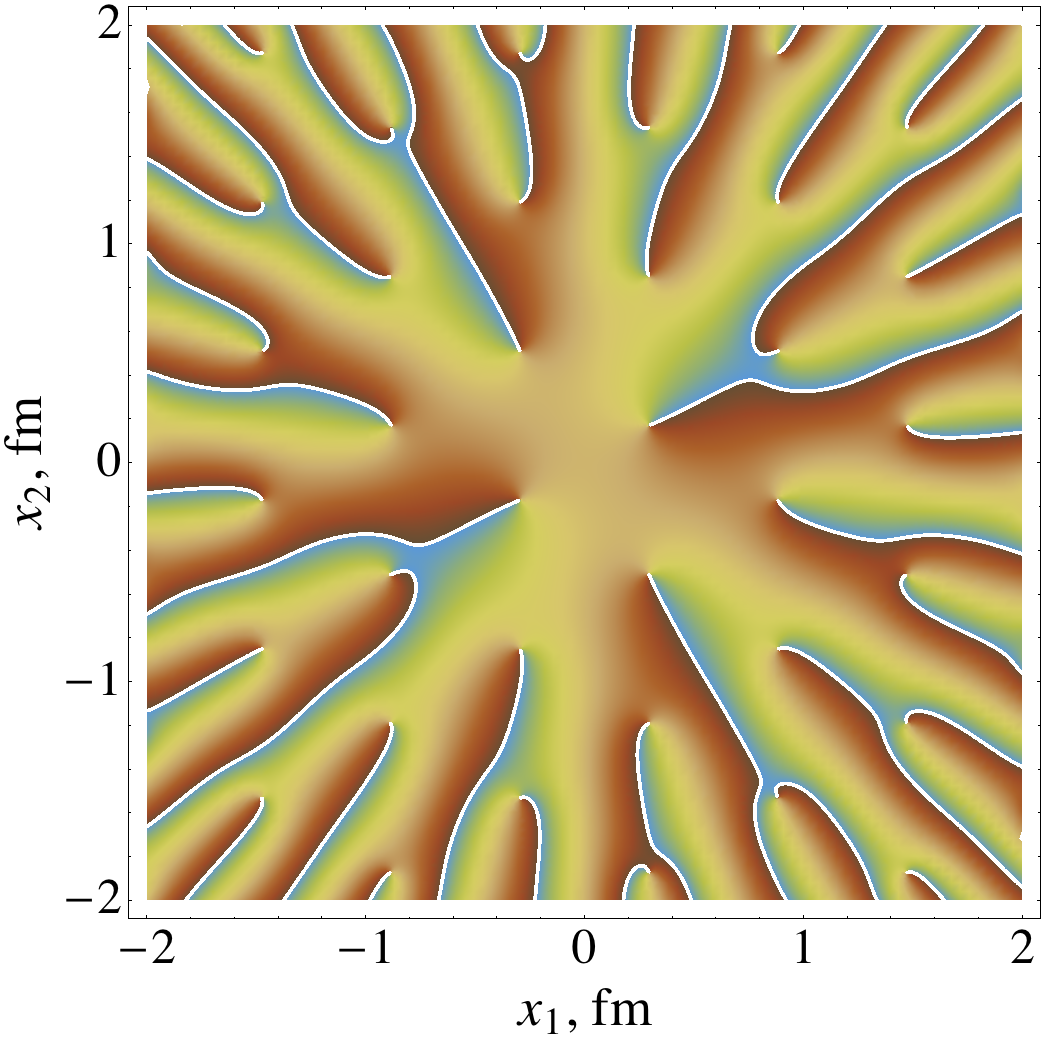} 
& \hskip 5mm
\includegraphics[width=50mm, angle=0]{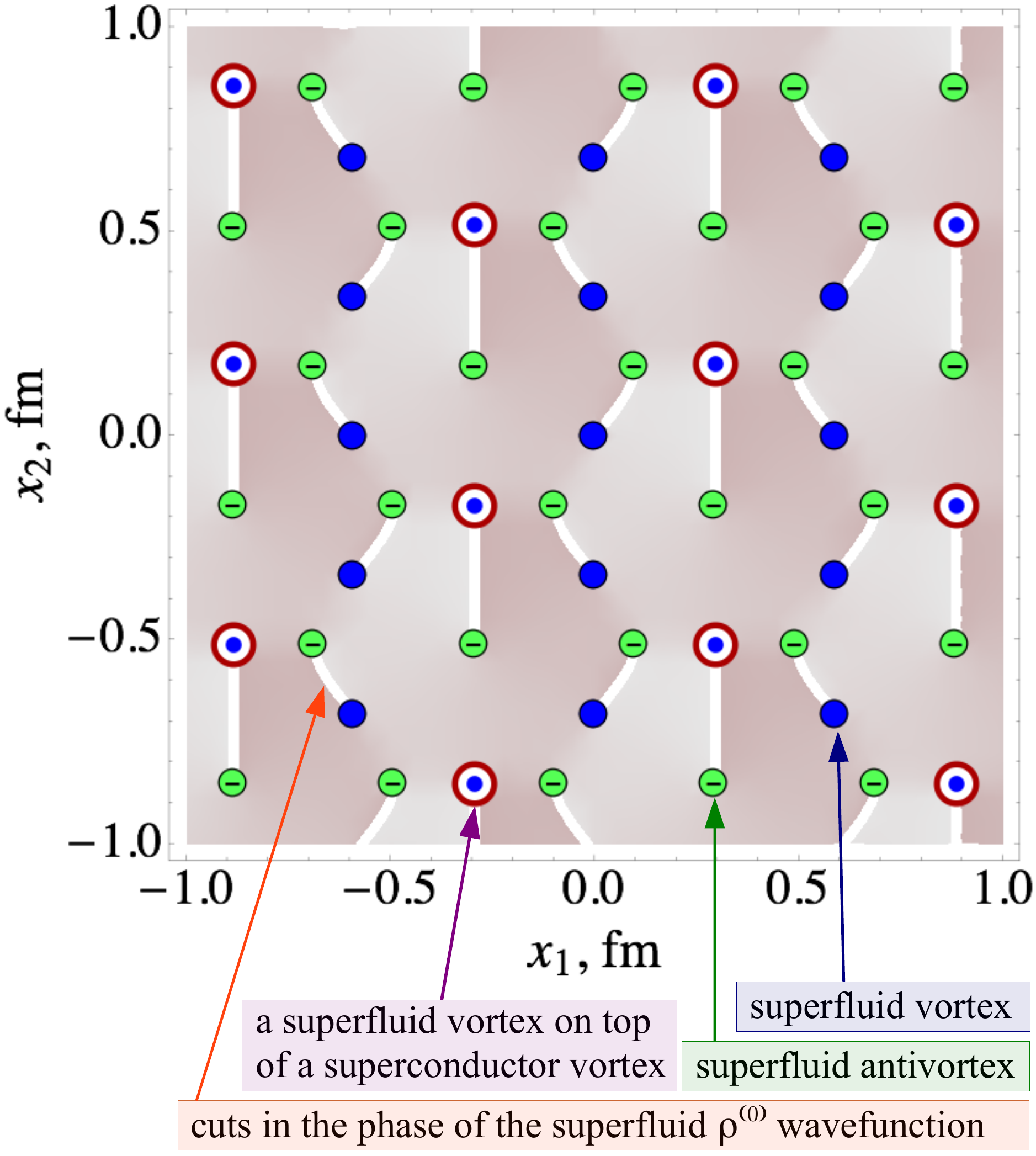}
\end{tabular}
\end{center}
\vskip -4mm
\caption{The same as in Fig.~\ref{fig:absolute} but for the phases for the charged ($\arg \rho$) and neutral ($\arg \rho^{(0)}$) condensates, respectively. The white lines represent the cuts in the corresponding phases. The endpoints of the white lines correspond to the superconductor and superfluid vortices, respectively. The plot on the right shows the positions of the superconductor and the superfluid vortices. Notice the difference in the coordinate scales of the plotted regions in Fig.~\ref{fig:absolute} and Fig.~\ref{fig:phases}  (Ref.~[29]).}
\label{fig:phases}
\end{figure}

\begin{figure}
\begin{center}
\begin{tabular}{ll}
\hskip 8mm\includegraphics[width=50mm, angle=0]{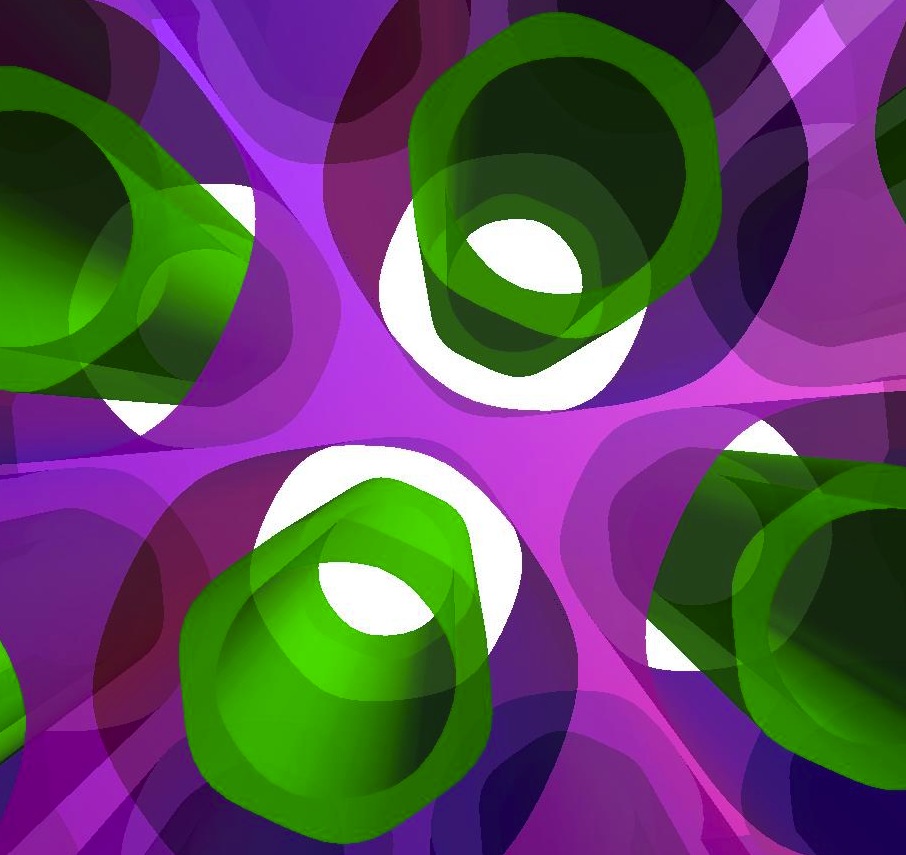} 
& \hskip 7mm
\includegraphics[width=46.5mm, angle=0]{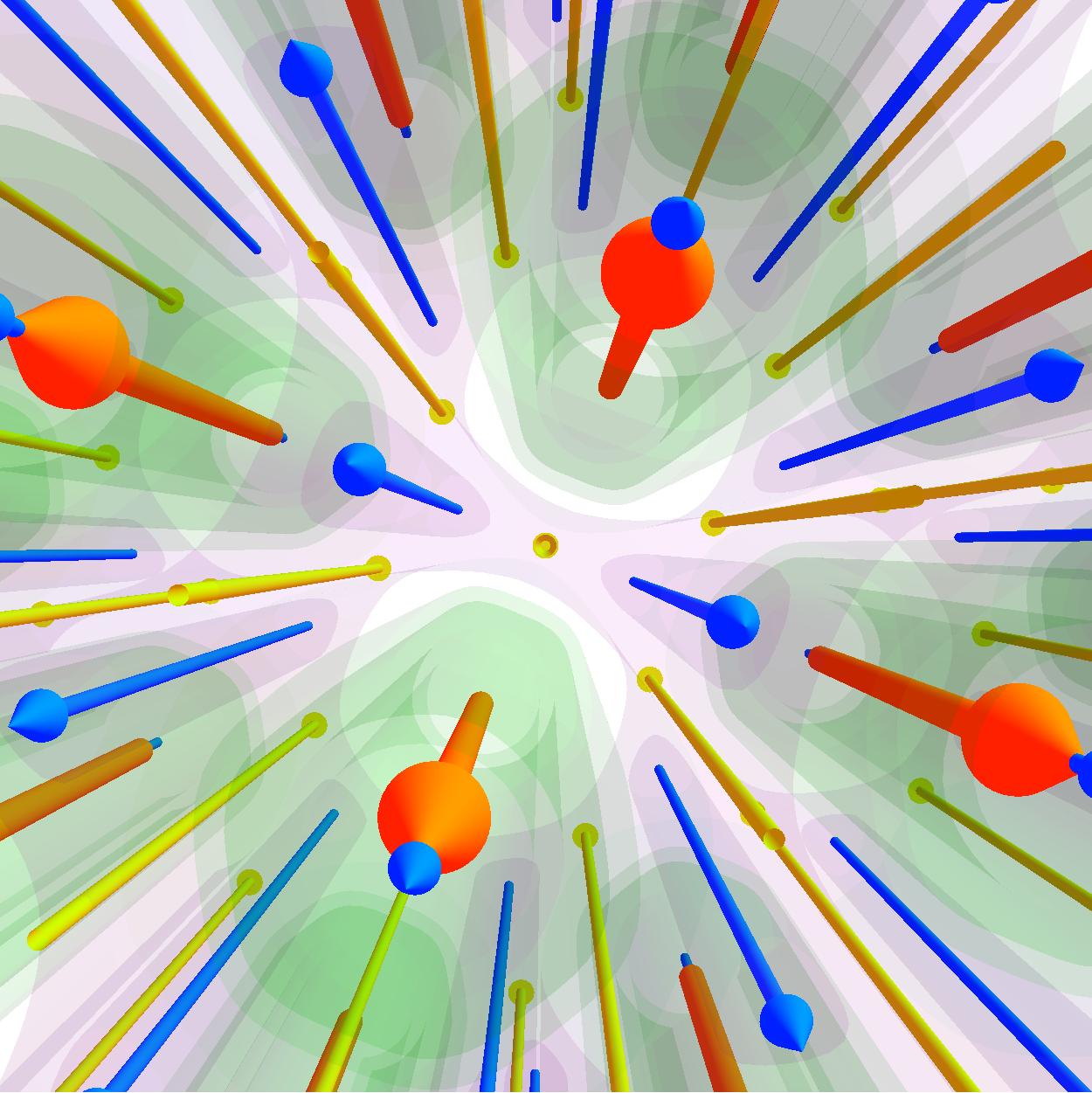}
\end{tabular}
\end{center}
\vskip -4mm
\caption{A $3d$ visualization of the ground state. (left) The honeycomblike structure of charged and neutral condensates. (right) The vortex lattice superimposed on the condensates. The magnetic field is directed out of the plane of the drawing.  The charged and neutral condensates approach their maxima in the violet and green regions, respectively.  The superconductor  and superfluid vortices are shown by, respectively, the large red and small blue arrows directed out of the plane. The superfluid antivortices are shown by the small yellow arrows directed into the plane.}
\label{fig:3d}
\end{figure}

The absolute values of the charged and neutral $\rho$--meson condensates are vanishing in an infinite set of isolated points. The corresponding phases, Figure~\ref{fig:phases}(right) and (left), wind around these points indicating that these points are nothing but the coordinates of the superconductor\cite{Chernodub:2010qx} and superfluid\cite{ref:Jos} vortices, respectively [Figure~\ref{fig:phases}(right)]. A $3d$ visualization of the condensates and vortices is shown in Figure~\ref{fig:3d}.

\vskip -2mm

\section{Instead of Conclusion}
The discussed effect is very unusual: an empty space becomes a superconductor if subjected to a strong enough background magnetic field. It is amazing that the magnetic field induces the superconductivity instead of destroying it. This effect can be considered as a (``magnetic'') analogue of the Schwinger pair production because the emergent superconductivity connects -- as in the Schwinger effect -- the real world with the virtual world: the strong magnetic field makes the real superconducting condensate out of the virtual particles.

\end{document}